\documentclass[12pt]{iopart}
\usepackage{graphicx}
\usepackage{iopams}
\begin{document}
\title{Pressure  effects on the electron-doped high $T_c$ superconductor BaFe$_{2-x}$Co$_{x}$As$_{2}$}
\author{K. Ahilan$^1$, J. Balasubramaniam$^1$,  F. L. Ning$^1$, T. Imai $^{1,2}$, A. S.
Sefat$^3$, R. Jin$^3$, M. A. McGuire$^3$, B. C. Sales $^3$, and D.
Mandrus$^3$}
\address{$^1$ Department of Physics and Astronomy, McMaster
University, Hamilton, Ontario L8S4M1,Canada}
\address{$^2$ Canadian Institute for Advanced Research, Toronto, Ontario M5G1Z8}
\address{$^3$ Materials Science and Technology Division, Oak Ridge, National Laboratory, TN 37831, USA}
\ead{imai@mcmaster.ca}
\begin{abstract}
Application of  pressures or electron-doping through Co substitution into Fe sites transforms the itinerant antiferromagnet BaFe$_{2}$As$_{2}$ into a superconductor with the $T_c$ exceeding 20~K.  We carried out systematic transport measurements of BaFe$_{2-x}$Co$_{x}$As$_{2}$ superconductors in  pressures up to 2.5~GPa, and elucidate the interplay between the effects of electron-doping and  pressures.  For the underdoped sample with nominal composition $x = 0.08$, application of  pressure strongly suppresses a magnetic instability while enhancing $T_{c}$ by nearly a factor of two from 11~K to 21~K.  In contrast, the optimally doped $x=0.20$ sample shows very little enhancement of $T_{c} = 22$~K under applied pressure.  Our results strongly suggest that the proximity to a magnetic instability is the key to the mechanism of superconductivity in iron-pnictides.

\end{abstract}
\pacs{74.25.Fy, 74.62.Fj} \submitto{\JPCM} \maketitle

The recent discovery of superconductivity at 26~K in the carrier-doped quaternary iron oxy-pnictide LaFeAsO attracted tremendous interest among the scientific community around the world \cite{kamihara}.
More recently, the ternary iron-pnictide family RFe$_{2}$As$_{2}$ (R= Ba, Sr, Ca) with the tetragonal ThCr$_{2}$Si$_{2}$ structure also has been shown to become a high $T_c$ superconductor upon doping \cite{rotter1}.  The common structural building blocks of  the LaFeAsO and RFe$_{2}$As$_{2}$ families are the square-lattice Fe sheets hybridized with As layers, hence there is no doubt that superconductivity takes place in the FeAs layers.  However, the mechanism of superconductivity is highly controversial.  The RFe$_{2}$As$_{2}$ family is a promising model system for a detailed investigation of the physical properties of the new iron-based high $T_c$ superconductors, because large single crystals are available.   BaFe$_{2}$As$_{2}$ is an itinerant antiferromagnet, and exhibits simultaneous structural (tetragonal to orthorhombic) and magnetic phase transitions  at $T_{SDW} \sim 140$~K \cite{rotter2,huang}.  Holes doped into the Fe sheets through substitution of K$^{+}$ ions into the Ba$^{2+}$ sites suppress the magnetic instability, and induce superconductivity with $T_{c}$ as high as 38 K in Ba$_{1-x}$K$_{x}$Fe$_{2}$As$_{2}$ with $x=0.45$ \cite{rotter1}.
This strategy of transforming an antiferromagnetic parent phase into a superconductor by doping from the Ba$_{1-x}$K$_{x}$ charge-reservoir layers is analogous to the recipe used in high $T_c$ copper-oxide superconductors.

However, two recent discoveries show that there are fundamental differences between cuprates and iron-pnictides.  First, one can induce superconductivity in the Fe sheets by substituting Co \cite{sefat1,sefat2,german} or Ni \cite{li}atoms directly into the Fe sites.  Co (Ni) atoms have one (two) more electron(s) than Fe atoms, and they donate these extra electrons to FeAs layers as superconducting carriers.  This phenomenon is in remarkable contrast with the case of cuprates; doping non-magnetic Zn$^{2+}$ ions with one extra electron into Cu sites induces local magnetic moments and destroys superconductivity \cite{fukuzumi,nachumi}.   Our $^{59}$Co and $^{75}$As NMR measurements in  BaFe$_{1.8}$Co$_{0.2}$As$_{2}$ showed no evidence for induced magnetic moments \cite{ning}.  Second, one can also suppress the itinerant antiferromagnetism of the parent phase RFe$_{2}$As$_{2}$ and induce superconductivity without doping by applying pressures \cite{canfield1,park,alireza}.  In cuprates, application of pressures does not transform the Mott-insulating state into a metallic state.

To date, the mechanism of carrier {\it or} pressure-induced superconductivity in RFe$_{2}$As$_{2}$ is not understood well, calling for further investigations.  In this report, we shed new light on this fascinating problem by exploring the interplay between carrier doping {\it and} pressures through systematic measurements of the electrical resistivity $\rho$ in Co-doped BaFe$_{2-x}$Co$_{x}$As$_{2}$ single crystals in pressures up to $P=2.5$~GPa.  We choose two nominal concentrations for the present study: the underdoped BaFe$_{1.92}$Co$_{0.08}$As$_{2}$ ($T_{c}$ $\sim 11$~ K) and optimally doped BaFe$_{1.8}$Co$_{0.2}$As$_{2}$ ($T_{c} \sim 22$~K).  Notice that the Co dopants replace nominally 4\% and 10\% of Fe sites in $x=0.08$ and $x=0.20$ samples, respectively.  We will demonstrate that application of pressure drives $x=0.08$ away from a magnetic instability, while enhancing $T_c$ by nearly a factor of two.  The highest $T_{c} \sim 21$~K observed for $x=0.08$ in 2.5~GPa is almost as high as $T_c \sim 22$~K of $x=0.20$, and indicates that the underdoped Fe sheets are as susceptible to pressures as undoped RFe$_{2}$As$_{2}$.  In contrast, we show that $T_c$ of the optimally doped $x=0.20$ barely changes with pressure.  Given that application of pressures also suppresses the magnetic instability in $x=0$ \cite{fukazawa} and $x=0.08$, we argue that the proximity to a magnetic critical point plays a crucial role in the superconducting mechanism.

Single crystals of Co-doped BaFe$_{2}$As$_{2}$ were grown out of FeAs flux, and characterized by X-ray diffraction and electron-probe micro-analysis. Further details can be found in \cite{sefat2}.  We cut small piece of crystals with typical dimensions 1 mm $\times$ 1mm $\times$ 0.15 mm from a larger piece for the electrical resistivity measurements in applied pressure. Sample contacts were made using Epotek silver epoxy for conventional four-probe ac-measurements. We employed a compact double-layered hydrostatic cylindrical clamp cell with a BeCu outer jacket and a NiCrAl alloy inner core to achieve pressures up to 2.5~GPa. Daphene oil 7373 and 99.99$\%$ purity Sn were used as a pressure transmitting medium and a pressure calibrating gauge, respectively.  In order to achieve high precision, we used a highly flexible home-built  ac-measurement system.  We stabilized temperature of the high pressure cell for every data point, instead of continuously sweeping the temperature. We confirmed the consistency of our results by conducting both cooling and warming measurements for every run.

We begin our discussions with the transport properties of optimally electron-doped BaFe$_{1.8}$Co$_{0.2}$As$_{2}$.  In Fig.1, we present the temperature dependence of the in-plane (ab-plane) resistivity $\rho_{ab}$ measured for $x=0.20$ at various hydrostatic pressures. $\rho_{ab}$ decreases monotonically with temperature, and we can fit the temperature dependence with a simple power law, $\rho_{ab} = A +B T^{n}$, with $n = 1.33\pm$0.05 in a broad range of temperatures from $\sim 300$~K down to $\sim 100$~K.  Below $\sim100$~K, the resistivity shows T-linear behavior $\rho_{ab} \sim T$ ($n=1$)  down to $T_{c}$ \cite{sefat2}.  Our recent NMR measurements in the same crystal showed that spin susceptibility $\chi_{spin}$ also undergoes a crossover between two different regimes across $\sim100$~K; $\chi_{spin}$ decreases monotonically with a pseudo-gap $\Delta_{PG}/k_B \sim 560\pm 150$~K, then almost levels off below $\sim100$~K \cite{ning}.  This means that low energy spin excitations are abundant in the high temperature regime above $\sim 100$~K, and the power $n = 1.33 \cong 4/3$ may be associated with the scattering of electrons by these excitations.  We emphasize that although the temperature independence of  $\chi_{spin}$ alone might suggest that BaFe$_{1.8}$Co$_{0.2}$As$_{2}$ is a Fermi-liquid system below $\sim100$~K, the T-linear behavior of $\rho$ indicates the contrary.  We recall that Fermi liquid systems would satisfy $\rho_{ab} \sim T^{2}$.  Instead, the T-linear behaviour is the benchmark of strongly correlated phenomena in copper-oxide high $T_c$ cuprates.  For example, optimally doped La$_{2-x}$Sr$_{x}$CuO$_{4}$ exhibits $\rho_{ab} \sim T$ for a wide range of temperatures from T$_{c}$ to as high as $\sim 1000$~K \cite{nakamura}.

Upon applying hydrostatic pressure, the magnitude of $\rho_{ab}$ in $x=0.20$  decreases substantially in the entire temperature range above $T_c$, but the temperature dependence shows very little change.  To illustrate this point, in Fig.2 we plot $\rho_{ab}$ measured for various pressure values by normalizing the magnitude at 290 K to the result obtained for 2.5~GPa.  All the data set collapse on top of each other except near $T_c$.  The pressure effect on $T_{c}$ is also weak.  $T_{c}$  increases only by $\sim 1$~K from $T_{c}\sim$ 22~K at $P=0$ to $T_{c} \sim 23$~K at $P=1.25$~GPa, then almost saturates, as summarized in Fig. 3.  The pressure coefficient below 1.25~GPa, $dT_{c}/dP \sim 0.65$~ K/GPa is by an order of magnitude smaller than that of the undoped antiferromagnetic phase BaFe$_{2}$As$_{2}$ \cite{alireza}.  In the latter case, $T_{c} = 0$~K increases quickly to 29~K in a narrower pressure range from 2.5~GPa to 4~GPa as shown in Fig.3.  These contrasting results indicate that once the magnetic instability is completely suppressed by 10\% Co substitution into Fe sites, application of pressures has little effect on $T_c$ in the optimally electron doped system.  It remains to be seen if applying higher pressures than 2.5~GPa would further enhance $T_c$, or even suppress $T_c$ to show a "dome shape" in the $P$ vs. $T_{c}$ phase diagram.   However, a naive extrapolation of $T_c$ in BaFe$_{1.8}$Co$_{0.2}$As$_{2}$  to higher pressure values does not seem to achieve as high a $T_{c} \sim 29$~K as in the undoped parent phase.  One obvious possibility is that the disorder induced by Co substitution into Fe limits the highest $T_c$, but this scenario seems unlikely.  The normal state residual resistivity at 2.5~GPa estimated from the linear extrapolation to $T=0$ is $\rho_{ab}(T\rightarrow 0) \sim 0.06$~m$\Omega \cdot$cm, and this value is about the same as that of undoped BaFe$_{2}$As$_{2}$ at 13~GPa \cite{fukazawa}.  It is also interesting to notice that the sign of the pressure coefficient $dT_{c}/dP$ is different between the present case of electron doped BaFe$_{1.8}$Co$_{0.2}$As$_{2}$ and the hole doped Ba$_{0.55}$K$_{0.45}$Fe$_{2}$As$_{2}$. The latter has a negative coefficient, $dT_{c}/dP = - 1.5$~K/GPa \cite{canfield2}. This may be an indication that the Fermi energy $E_F$ of the undoped BaFe$_{2}$As$_{2}$ is located near a singularity of the density of states, and electron and hole-doping place $E_F$ on different slopes.

Next, we turn our attention to the pressure effects on the underdoped, Co 4\% doped BaFe$_{1.92}$Co$_{0.08}$As$_{2}$.  In Fig.4, we show $\rho_{ab}$ measured for $x=0.08$ at various hydrostatic pressures.  The temperature dependence above $\sim100$~K is qualitatively similar to that of the optimally doped  $x=0.2$.  An analogous power law fit to $\rho_{ab} = A +B T^{n}$ yields a somewhat larger value of the power $n = 1.60\pm$0.03 at ambient pressure as shown in Fig.2.  Unlike the optimally doped $x=0.20$, a monotonic decrease of $\rho_{ab}$ from $\sim 300$~K down to $\sim 95$~K is followed by a step-like increase near $\sim 70$~K.  The sharp dip centered around 66~K in the derivative presented in Fig.2b, $d\rho_{ab}/dT$, hints at the presence of a phase transition. To clarify the origin of the anomaly, we also carried out NMR measurements on the same batch of crystals \cite{ning2}. The dramatic NMR line broadening and a divergence of the nuclear spin-lattice relaxation rate 1/T$_{1}$ below $\lesssim 70$~K indicate that this anomaly is caused by a magnetic instability.  In what follows, we tentatively represent the magnetic anomaly temperature using the same symbol $T_{SDW}$ commonly used for the undoped BaFe$_{2}$As$_{2}$, but we need to keep in mind that the nature of the magnetic instability in the electron doped $x=0.08$ may be somewhat different.  In particular, whether a structural phase transition accompanies the magnetic instability at or near 70~K requires further investigation.  Regardless, it is worth pointing out that signatures of the critical slowing down observed above 70~K in our NMR data indicate that this magnetic instability is not strongly first order.  Since we don't observe any sharp paramagnetic NMR signals below $T_{SDW}$, we can also conclude that the magnetic instability at $T_{SDW}$ affects the entire Fe$_{1.92}$Co$_{0.08}$As$_{2}$ layers.  In passing, a similar step-like resistivity increase is also reported for  Co-doped LaFe$_{1-x}$Co$_{x}$AsO \cite{sefat1}, SrFe$_{2-x}$Co$_{x}$As$_{2}$ ($x = 0.1$ and 0.15) \cite{german}, and Ni-doped BaFe$_{1.95}$Ni$_{0.05}$As$_{2}$ \cite{li}.

With decreasing temperature, $x=0.08$ also exhibits a superconducting transition to a zero resistivity state below $T_c \sim 11$~K.  Given our NMR finding that the magnetic instability affects all sites in the Fe$_{1.92}$Co$_{0.08}$As$_{2}$ layers, we rule out the possibility that this superconducting transition is caused by macroscopic inhomogeneity in Co concentrations.  Since the magnetic instability precedes the superconducting transition, the superconducting state should be considered granular, in analogy with the case of underdoped high $T_c$ cuprates.  We may naturally understand the apparent coexistence of $T_{SDW}$ and $T_c$ based on the following simple argument.  The key point is that the superconducting coherence length $\xi \sim 3.5$~nm \cite{athena2} is very short in iron-pnictide superconductors.   Assume that each Co atom donates an electron to nearby Fe sites, destroys magnetism locally, and creates a superconducting island.  As a ballpark figure, one may assume that the diameter of the island is roughly equal to the in-plane superconducting coherence length, $\xi$, and involves more than ten Fe sites.  Then the Co doping level of 4\% is more than enough to achieve the classical percolation limit of 50\% for two-dimensions, and hence zero resistance.    This situation of the "reverse Swiss cheese model" is the opposite to the so-called "Swiss cheese model."  The latter describes the destruction of superconductivity in Zn doped high $T_c$ cuprates \cite{nachumi}.

We notice three striking features in the pressure effects in Fig.4.  First, with applying pressure, $T_{SDW}$ is markedly suppressed from $T_{SDW} \sim 66$~K to $\sim 52$~K in 2.5~GPa.  This strong pressure effect on $T_{SDW}$ is comparable to the case of the undoped BaFe$_{2}$As$_{2}$.  In the latter, $T_{SDW} \sim 130$~K is suppressed to $\sim 110$~K in 2.5~GPa \cite{fukazawa}; see the phase diagram in Fig.3 for comparison.  Second, as the magnetic instability is suppressed, the temperature dependence of $\rho_{ab}$ becomes increasingly similar to that of $x=0.2$.  We show the representative fits of $\rho_{ab}$ by a power-law in Fig.2, and summarize the pressure dependence of the power $n$ in Fig.5.  The power $n$ gradually decreases with increasing pressure from $n=1.6$ at 0~GPa to 1.4 at 2.5~GPa.  Fig.5 suggests that the power $n$ for $x=0.08$  approaches $n=1.33$ in higher pressures, which is the value observed for the optimumally doped $x=0.20$ sample.  It is tempting to interpret this as an indication that $n=4/3$ is the universal value in the metallic state above $\sim 100$~K of Co-doped superconducting BaFe$_{2-x}$Co$_{x}$As$_{2}$.  We caution, however, that a similar power-law fit of the resistivity data reported by by Fukazawa  et al. \cite{fukazawa}   for undoped BaFe$_{2}$As$_{2}$ in 13~GPa yields a smaller value, $n \sim 1.1$.  Third, application of pressures results in a very strong enhancement of $T_c$ from 11~K to 21~K in 2.5~GPa.  That is, one can enhance $T_c$ nearly by a factor of two.  We emphasize that the observed pressure coefficient, $dT_{c}/dP \sim 4$~ K/GPa is quite sizable, and nearly by an order of magnitude larger than 0.65~K/GPa of the optimally doped $x=0.20$ sample.  Equally interesting is our finding that $T_c$ of the $x=0.08$ sample shows no sign of saturation up to 2.5~GPa, unlike the case of $x=0.20$.  In fact, a naive extrapolation of $T_c$ in Fig.3 shows that $T_c$ of $x=0.08$ may exceed that of $x=0.20$ and approaches the maximum $T_c$ of $x=0$.  We recall that many earlier studies of pressure effects on $T_c$ in undoped or hole-doped iron-pnictides showed either a dome-like shape in the $T_c$ vs. $P$ phase diagram, or a linear decrease of $T_{c}$ at elevated pressures \cite{canfield1,park,alireza,canfield2,lorenz,takahashi,zocco,takeshita,kumar}.

To summarize, we demonstrated contrasting behaviors of pressure effects on superconductivity between optimally electron-doped BaFe$_{1.8}$Co$_{0.2}$As$_{2}$ and underdoped BaFe$_{1.92}$Co$_{0.08}$As$_{2}$.  We showed that once the magnetic instability is completely suppressed by electron doping, application of high pressures has very little effect on superconductivity and the qualitative behavior of the normal state resistivity.  On the other hand, the suppression of  a magnetic instability with pressure results in a large enhancement of $T_c$ by nearly a factor of two. This finding strongly suggests that magnetism and superconductivity are strongly correlated.  Needless to say, it does not necessarily mean that the glue of the superconducting Cooper pairs are magnetic fluctuations.  Instead, one can certainly view Fig.3 as evidence for competition between magnetism and superconductivity. In any event, our new results on underdoped $x=0.08$ establish a link between superconductivity and magnetism, and show  that fine tuning magnetic fluctuations with pressure is as effective as doping more carriers by additional Co doping.

\ack
T.I acknowledges helpful communications with A. J. Berlinsky and Y. Liu, and  financial support by NSERC, CFI and CIFAR.  Work at ORNL was supported by the Division of Materials and Engineering, Office of Basic Sciences.  A portion of this work was performed by Eugene P. Wigner Fellows at ORNL.\\

\clearpage
\begin{figure}
\centering
\caption{The temperature dependence of the in-plane resistivity $\rho_{ab}$ of a single crystal of BaFe$_{1.8}$Co$_{0.2}$As$_{2}$ at various pressures.  A straight line is drawn through the
ambient pressure data to show the T-linear behaviour up to $\sim 100$~K. The inset shows the expanded view near $T_c$.  We define $T_c$ as the midpoint of the transition.}
\label{figure:1}

\centering
\caption{Representative fits of the resistivity data above $\sim 100$~K with a power law, $\rho_{ab} = A + B T^{n}$, where $A$ and $B$ are constants.  For the optimally doped BaFe$_{1.8}$Co$_{0.2}$As$_{2}$, we normalized all the resistivity data at 290~K to the result obtained at 2.5~GPa.  For clarity, we show one data point of $\rho_{ab}$ for every four temperatures.}
\label{figure:2}

\centering
\caption{A summary of the $P-T$ phase diagram of magnetism and superconductivity in BaFe$_{2-x}$Co$_{x}$As$_{2}$ ($x=0.20$, 0.08, and 0).  The results of $T_c$ and $T_{SDW}$ for $x=0$ are by Alireza et al. \cite{alireza} and Fukazawa et al. \cite{fukazawa}, respectively.  All lines are a guide for the eyes.}
\label{figure:3}

\centering
\caption{Main panel: the temperature dependence of the in-plane resistivity $\rho_{ab}$ of a single crystal of BaFe$_{1.92}$Co$_{0.08}$As$_{2}$ at various pressures.  The vertical arrows show the magnetic ordering temperature at various pressures.  Inset: (a) the expanded view near $T_c$. (b) the derivative $d\rho_{ab}/dT$.  }
\label{figure:4}

\centering
\caption{The pressure dependence of the power $n$ for the power law fit of $\rho_{ab} = A + B T^{n}$ above $\sim 100$~K.  The straight line across the data points for $x=0.08$ underscores the pressure independence of $n$.  The solid curve for the $x=0.20$ data is a guide for the eyes.}
\label{figure:5}
\end{figure}

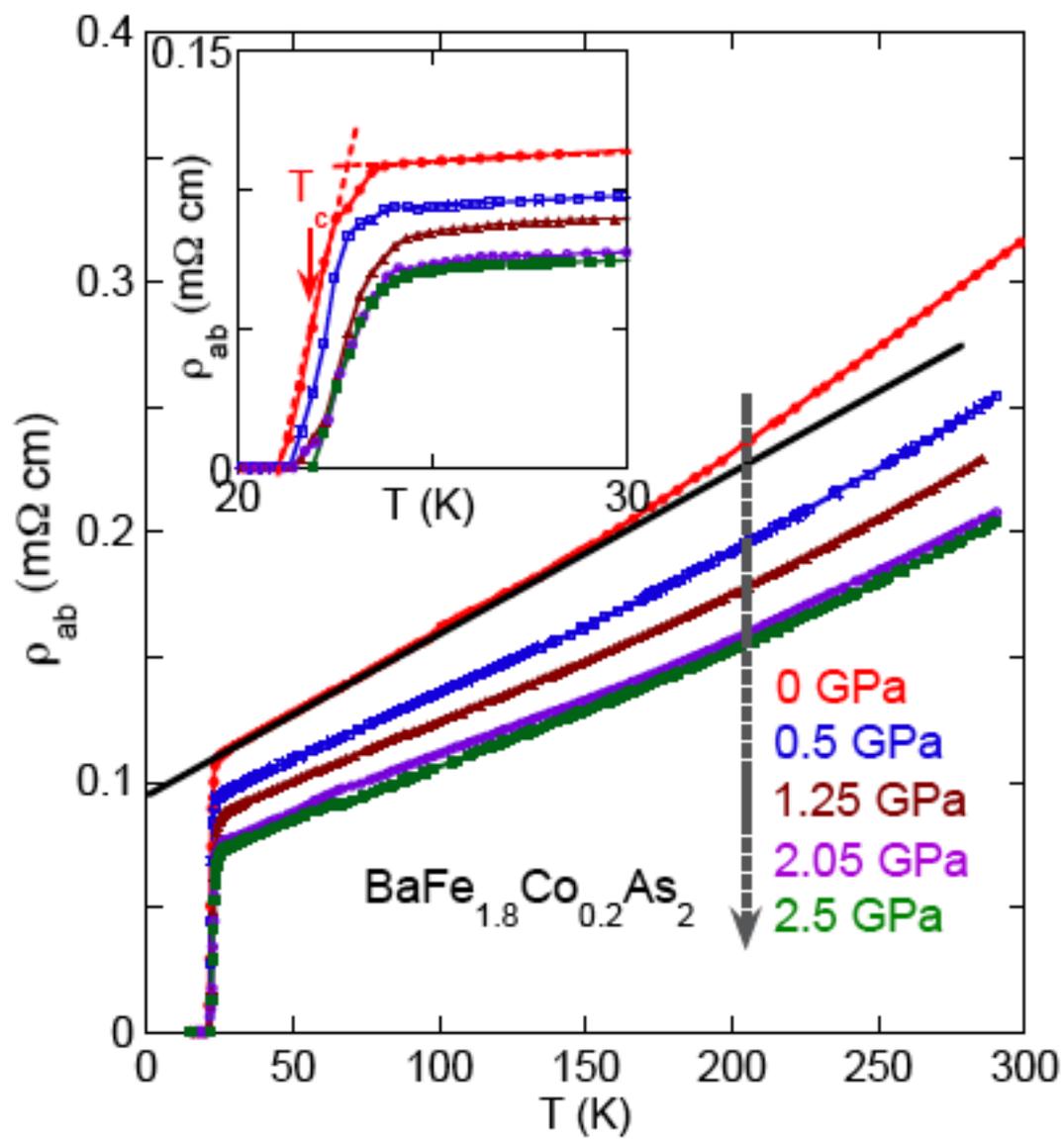
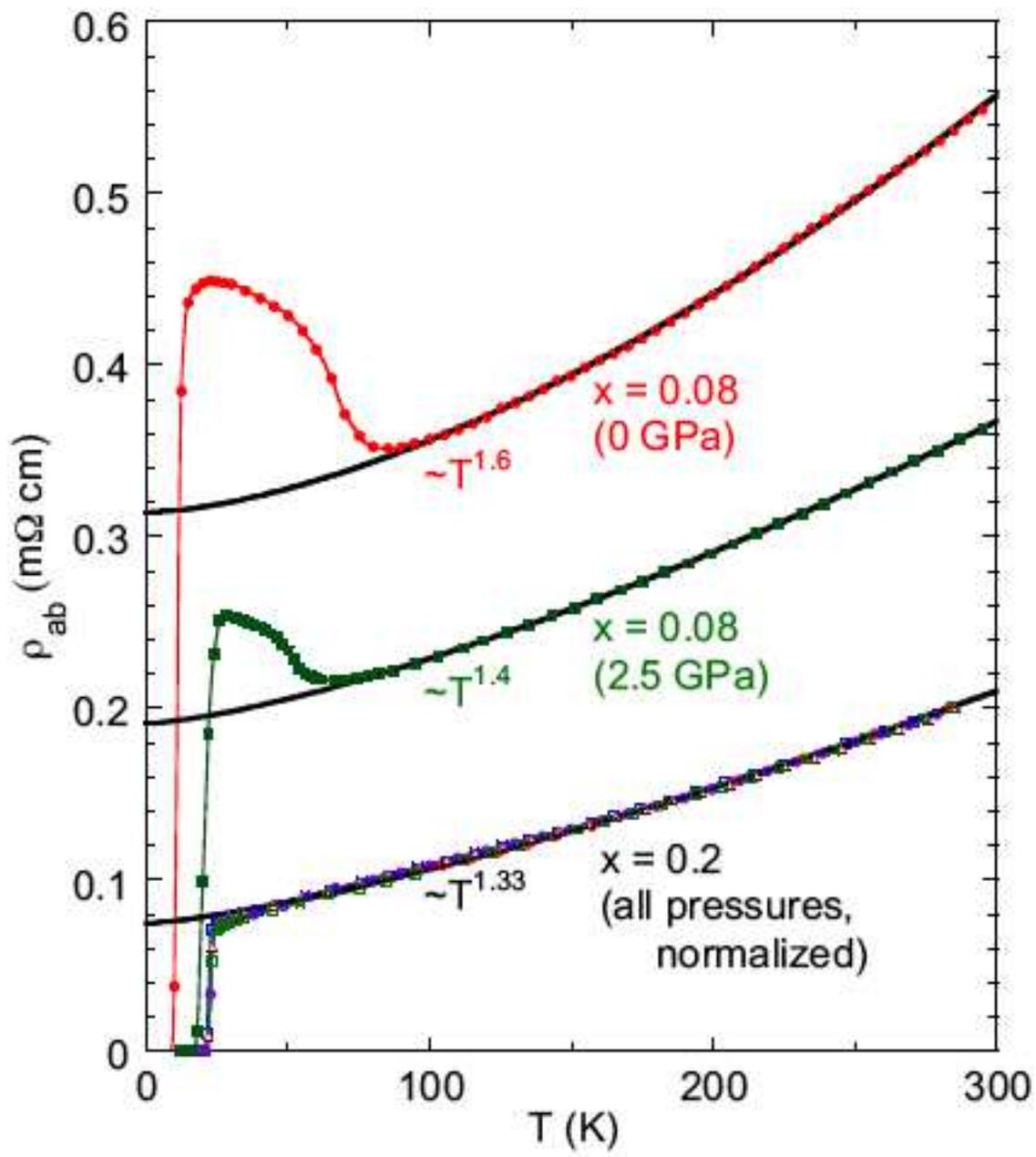
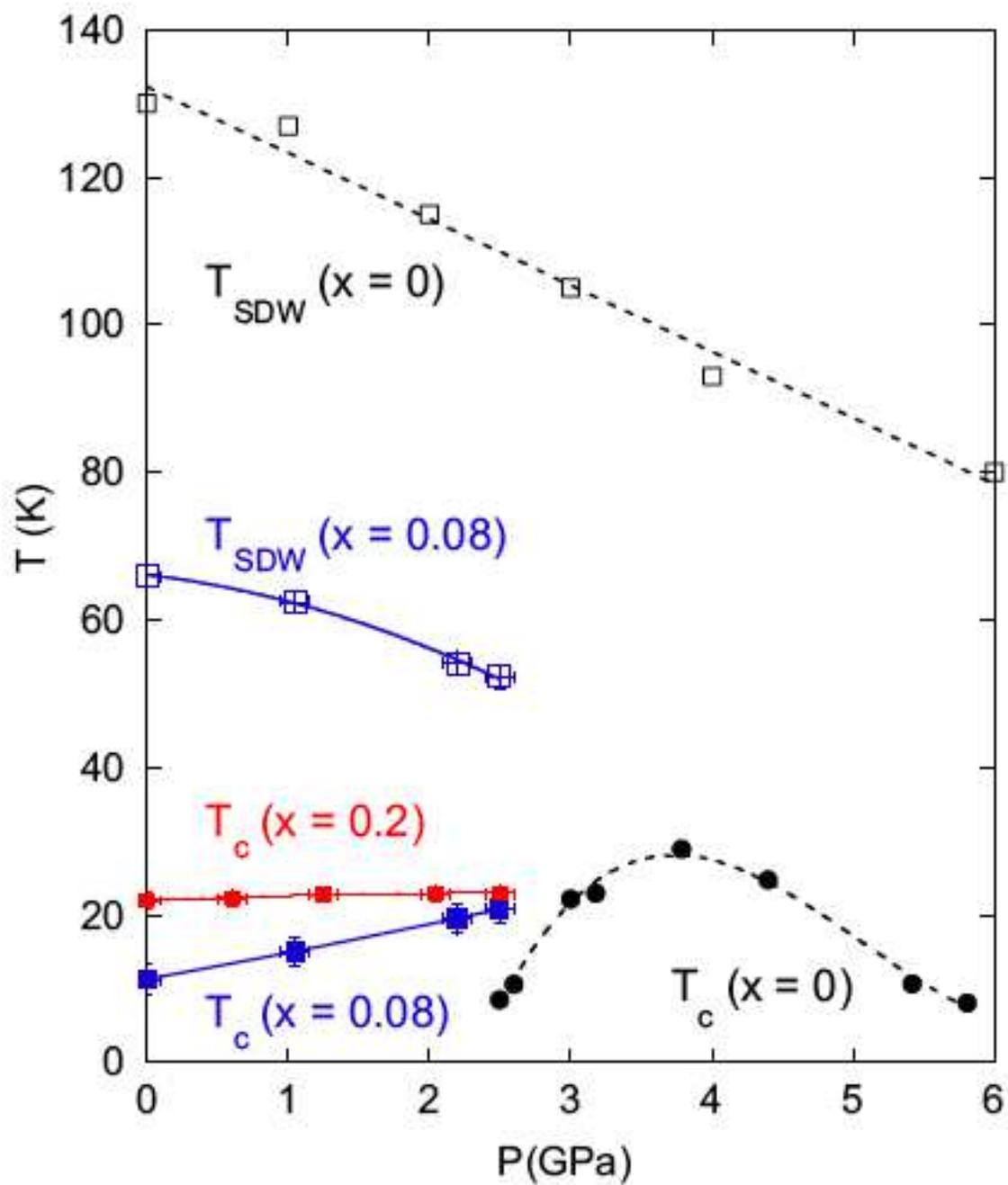
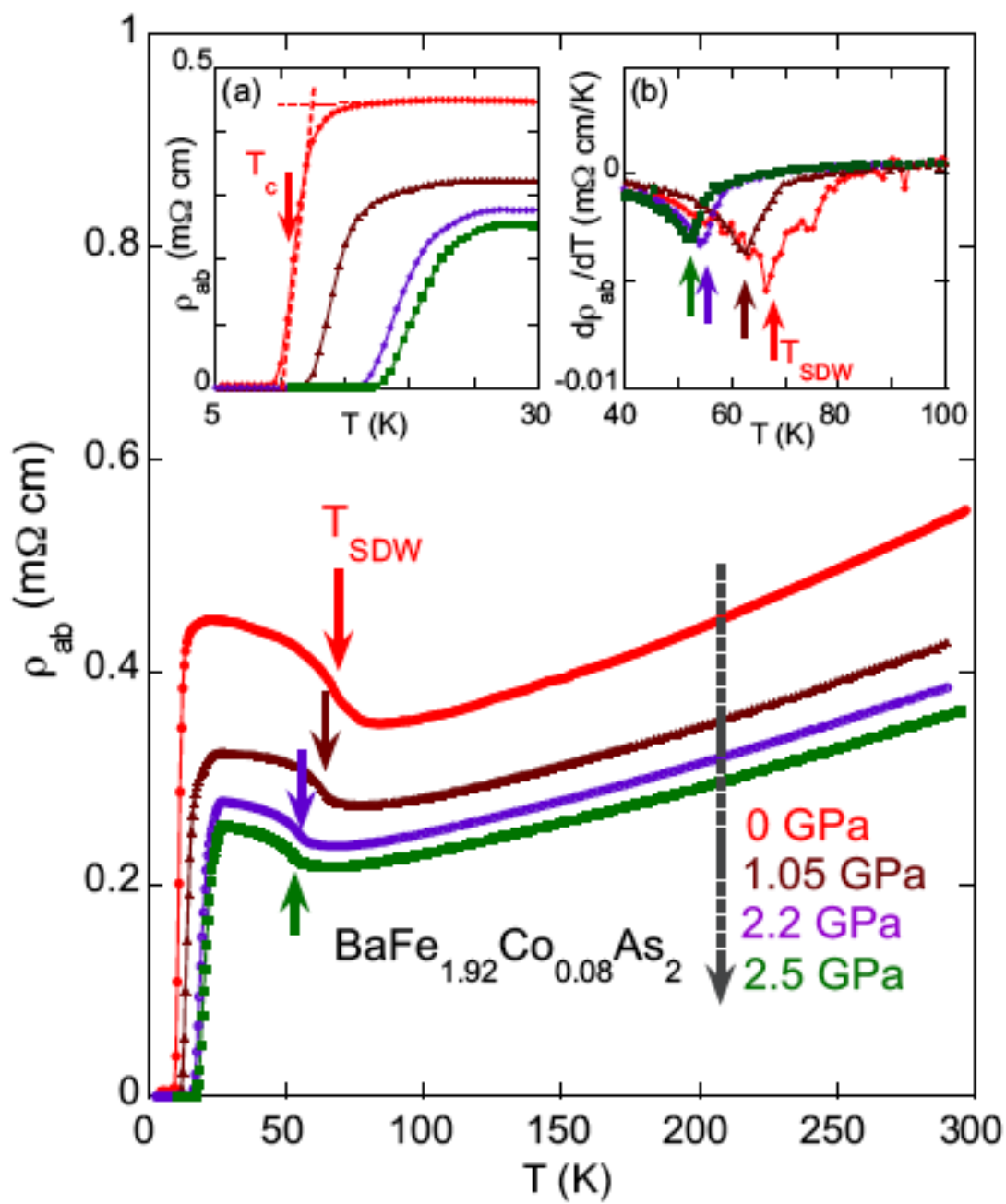
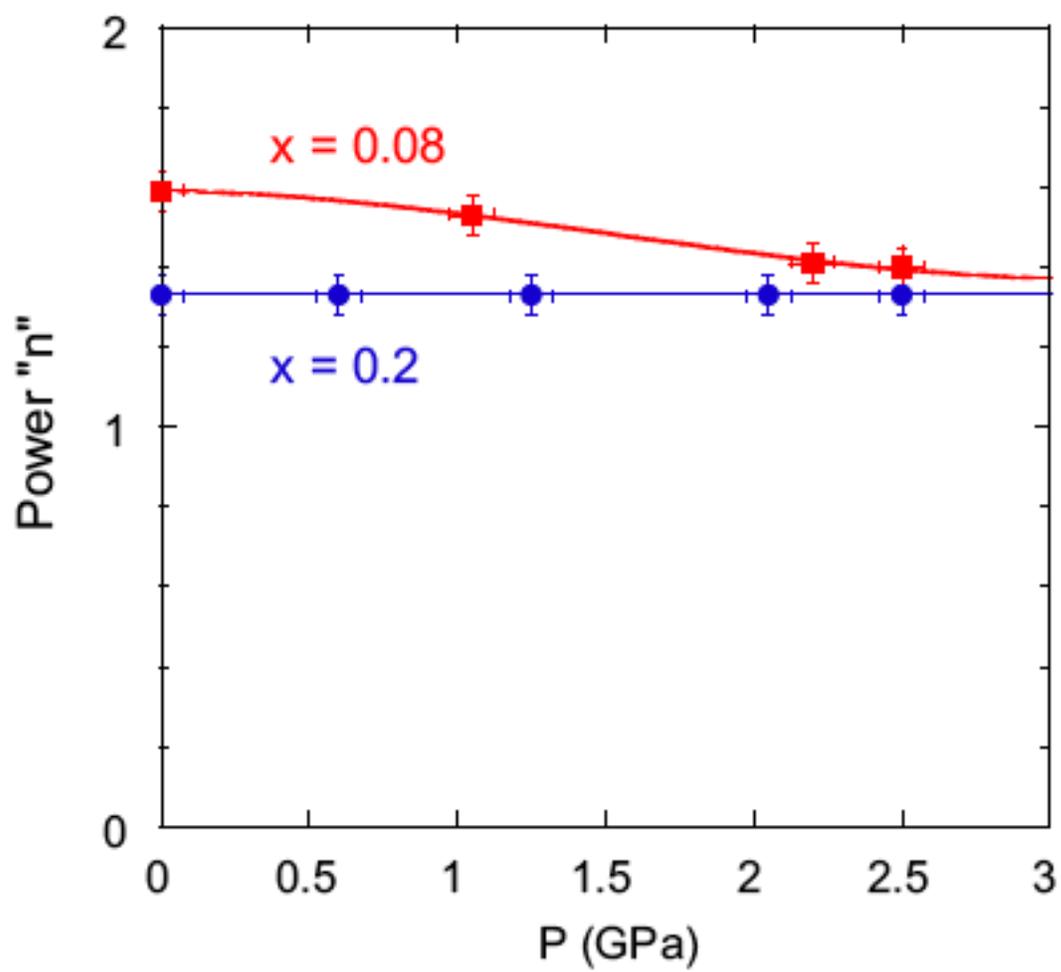
\end{document}